# Doubling the field of view and eliminating the replica overlap problem in common-path shearing quantitative phase imaging

Mikołaj Rogalski[1,*], Piotr Zdańkowski[1,**], Matyas Heto[1], Jolanta Mierzejewska[2], Małgorzata Lenarcik[3,4], Zhuoshi Li[5,6,7], Jiasong Sun[5,6,7], Chao Zuo[5,6,7], and Maciej Trusiak[1,8]

[1]Institute of Micromechanics and Photonics, Warsaw University of Technology, 8 Sw. A. Boboli St., 02-525 Warsaw, Poland

[2]Faculty of Chemistry, Warsaw University of Technology, Noakowskiego 3, 00-664 Warsaw, Poland

[3]Department of Pathology, Maria Sklodowska-Curie National Research Institute of Oncology, Warsaw, Poland

[4]Department of Gastroenterology, Hepatology and Clinical Oncology, Centre of Postgraduate Medical Education, Warsaw, Poland

[5]Smart Computational Imaging Laboratory (SCILab), Nanjing University of Science and Technology, Nanjing, Jiangsu 210094, China

[6]Smart Computational Imaging Research Institute (SCIRI), Nanjing University of Science and Technology, Nanjing, Jiangsu Province 210019, China

[7]Jiangsu Key Laboratory of Spectral Imaging and Intelligent Sense, Nanjing, Jiangsu 210094, China

[8]maciej.trusiak@pw.edu.pl

*mikolaj.rogalski@pw.edu.pl

**piotr.zdankowski@pw.edu.pl

**Abstract:** Quantitative phase imaging (QPI) enables label-free, high-contrast visualization of transparent specimens, but its common implementation in off-axis digital holographic microscopy (DHM) requires a separate reference beam, which increases system complexity and sensitivity to noise and vibrations. Common-path shearing DHMs overcome these drawbacks by eliminating the reference arm, yet they suffer from sheared object beam (replica) overlap, as both interfering sheared beams traverse the sample and generate superimposed phase images. This limits their use to sparse objects only. Here we introduce R2D-QPI, a method that numerically decouples object and replica fields of view through controlled shear scanning. The method analytically separates overlapped phase images and effectively doubles the imaged area, requiring only two measurements. We experimentally validate the approach on a phase resolution test target, yeast cells, and human thyroid tissue slices, demonstrating accurate reconstruction even in highly confluent samples with strong object–replica overlap. The results establish R2D-

QPI as a robust and versatile solution for common-path QPI, enabling wide-field, label-free phase imaging with minimal data acquisition and strong potential for applications in biological and medical microscopy.

# Introduction

Quantitative phase imaging (QPI), that synergizes phase contrast [1] and holography [2] principles, has revolutionized label-free microscopy by mapping optical path–length variations, which directly correspond to physical thickness profile and refractive-index (RI) characteristics of transparent specimens [3,4]. High-contrast phase measurements are fueling a growing number of applications in biomedicine [4–6], with examples in neuroscience [7], cancer research [8], organoid studies [9], and embryology [10], to name only a few. Coherent realization of QPI traditionally involves off-axis digital holographic microscopy (DHM) [11–13], which generates a hologram via interference of inclined beams [14] coming from object and reference arms of a Mach-Zehnder type interferometric configuration. Complex fields are retrieved via Fourier-transform algorithms [15], which, when combined with multi-directional oblique illumination, open the path to label-free super-resolution imaging, surpassing the diffraction limit and enabling nanoscale visualization of transparent specimens [16]. Although off-axis DHM is highly capable, its need for high spatial and temporal coherence makes it prone to laser speckle noise [17,18] and parasitic interferences, generally degrading image quality [11,12]. On top of that, off-axis DHM require very stable working conditions, as any environmental disturbances, such as vibrations, cause fringe instability and contrast loss, which reduces the final quality of the phase reconstruction.

Shearing interferometry mitigates the "two-arm" instability drawbacks by letting the object wave interfere with its own laterally shifted (sheared) replica [19,20]. An object-free region acts as the reference wave and allows for efficient hologram formation for further quantitative phase extraction [21,22]. Shearing is induced to a sample beam typically by shearing plate- [23–26], diffraction grating- [27–32], Wollaston prism- [33], beam displacer- [34,35], Michelson- [36–40] or Sagnac- [41] based interferometric modules. Common-path configuration of shearing interferometry [42,43], where object and reference waves travel along nearly identical optical paths, is favorable due to environmental perturbations cancelling out in the interference signal. Moreover, due to the near-equal length of optical paths of the interfering beams, low-coherence sources such as LEDs or filtered supercontinuum lasers can be used to further suppress speckle noise [44] while maintaining high phase sensitivity [45–48].

Despite their simplicity and compatibility with standard microscopes, common-path shearing systems, in a total-shear regime, limit the field of view (e.g., to 1/2 [29–31] or 1/3 [29,31,39]) and suffer a fundamental limitation in dense or confluent samples due to a lack of object-free reference areas. Hence, for specimens such as tissue sections or crowded cell cultures, the sample beam replicas (+1 and –1 diffraction orders in case of grating-based systems) overlap across the entire available field of view (FOV). In this regime, single-frame phase retrieval breaks down, and the usable FOV is significantly limited, if not entirely gone. One solution is to image only sparse objects and force the object placement so that one of the beam replicas pass through sample-free area; however, it narrows down the versatility of this group of QPI techniques application-wise.

Another methods of compensating this problem is the use of multiple images with variable shears and numerically remove the object replica from the FOV. One of the proposed solutions [22] used alternating projections algorithm to decouple replicas, but it required low-pass filtering and rotation of both object beams so that each replica contains negative and positive shears, At minimum, 24 images were required for proper replica removal, limiting the throughput and resolution of the method. Another work [49–51] proposed Cepstrum-based interference microscopy (CIM) to remove replicas, however it requires a precise, integer 1 pixel shifting of replicas in both x and y directions. CIM likewise employs high-pass filtering, attenuating low-frequency information.

Another group of common-path shearing interferometry techniques are gradient phase microscopy methods, which offer alternative routes to phase contrast with very small shear values [52–58]. These methods excel at high-contrast gradient imaging but necessitate numerical integration that amplifies low-frequency noise [59] and often require specialized hardware such as metamaterials [54–56] or multiple exposures. Thus, direct phase imaging for highly confluent samples in shearing common-path scenarios remains an unresolved issue in optical imaging community.

To address these challenges, we introduce R2D-QPI (Replica-Removed Doubled-FOV Quantitative Phase Imaging), a general framework that combines shear-scanning common-path interferometry with a dedicated Multi-shear Replica Removal (MRR) algorithm. The approach restores quantitative phase from fully overlapping beams by analytically disentangling object and replica contributions, while simultaneously extending the usable field of view. In this work, we validate the method in the grating-based common-path system. By axially translating the diffraction grating in small increments (tens to hundreds of microns), we induce controlled variations in lateral shear for the +1 and –1 orders. Recording a short sequence of interferograms (>= 2 frames), our novel algorithm exploits these shear differences to decouple conjugate replicas and recover +1 and –1 quantitative phase fields independently – without any gradient generation or integration. Comparing to standard total-shear realizations, our method is effectively doubling the FOV of QPI system (instead of reducing it to 1/2 or 1/3 of FOV), which is an additional advantage in high-throughput examination. Moreover, the proposed method has no limitations on the lack of object overlap for +1 and –1 FOVs, allowing to reconstruct even a very dense and highly confluent samples. Notably, the method is not limited to the grating-based system, it can be applied to any common-path configuration, providing it can introduce tunable shears. We validate R2D-QPI on diffraction-limited phase targets under coherent and low-coherence illumination, on highly confluent yeast layers where standard total-shear methods fail, and on human tissue sections, demonstrating true phase imaging with doubled FOV while preserving the simplicity, robustness, and low-coherence compatibility of common-path grating interferometry.

## Methods

### Experimental system

The experimental setup is based on our previously reported polarization-grating-aided common-path QPI and ODT system [48], and is schematically illustrated in Figure 1(a). The sample is illuminated by a collimated beam that is either coherent or partially coherent. The beam passes through a microscope objective (MO, Nikon Plan Apo 20×/0.75 or Zeiss 5×/0.12) and a tube lens (TL), forming the object wave. This wavefront is subsequently directed into the shearing module, shown in detail in Fig. 1(b), and finally recorded at the camera (Daheng MER2-301-125U3M).

The shearing module consists of two identical polarization diffraction gratings (PG1 and PG2) and a polarizer. Under linearly polarized illumination, each grating produces only the ±1st diffraction orders, which are circularly polarized (CP) with opposite handedness. PG1 splits the incident object beam into two beams propagating at angles $\pm\theta_1$, where the diffraction angle is given by $\theta_1 = \sin^{-1}(\lambda/\Gamma)$ with $\lambda$ denoting the wavelength and $\Gamma$ the grating period ($\Gamma = 6.29$ μm for our gratings). These two beams then encounter PG2, which reverses their CP handedness and deflects them further by an angle $\theta_2 = \sin^{-1}(\cos(\beta) \cdot \lambda/\Gamma)$ where $\beta$ is the relative rotation angle between the line orientations of PG1 and PG2. When the gratings are aligned ($\beta = 0$), the deflection angles satisfy $\theta_2 = \theta_1$, and the two beams emerging from PG2 become parallel. However, when PG2 is rotated ($\beta > 0$), the output beams diverge at an angle $\alpha = 2(\theta_1 - \theta_2)$. Additionally, the beams then undergo a mutual rotation around the optical axis that depends on $\beta$ – this rotation is omitted in Fig. 1 for clarity.

After PG2, both beams pass through a polarizer that restores their common linear polarization before reaching the camera plane. In the overlap region of the two beams (Fig. 1(c)), an interference pattern is formed with a carrier frequency determined by the angle $\alpha$. By adjusting $\alpha$ via rotation of PG2, we can easily control the spatial frequency of the fringes. In this work, $\alpha$ was chosen to produce dense carrier fringes (Fig. 1(d)), allowing clean separation of the interference terms in Fourier space (Fig. 1(e)) and subsequent retrieval of the phase image (however with overlapped FOVs coming from ±1st orders) using the Fourier transform method [15].

The shear between the two interfering beams ($\tau$) depends primarily on the distance between PG1 and PG2 ($z_{PG}$), as well as on the angle $\alpha$ and the PG2–camera distance. In our configuration, $\tau$ was set to exceed the camera diagonal, such that the camera records two independent FOVs corresponding to the interfering beams. This effectively doubles the imaged area. The method, however, can also be operated with smaller $\tau$ values, in which the same object region appears in both ±1st diffraction order FOVs, albeit without the FOV doubling effect.

To separate the overlapped +1 and –1 order phase contributions, we record multiple interferograms while slightly varying $\tau$. This is achieved by axially translating PG2 by a distance $\delta$ (typically 0.05–0.5 mm), which corresponds to a shear change of $2\Delta r$ in Fig. 1(b) (typically by several camera pixels). The PG2 translation is implemented using a motorized linear stage, enabling automatic and repeatable acquisition. Notably, precise control of $\delta$ is not required: the reconstruction algorithm only needs the corresponding $2\Delta r$ values with pixel-level accuracy, which can be determined directly from the observed object displacement in the recorded data if $\delta$ is not perfectly calibrated or the $\delta \to 2\Delta r$ relation is unknown. The resulting interferogram sequence, along with the estimated $2\Delta r$ shear values, is then used as input for the reconstruction algorithm described in the following section.

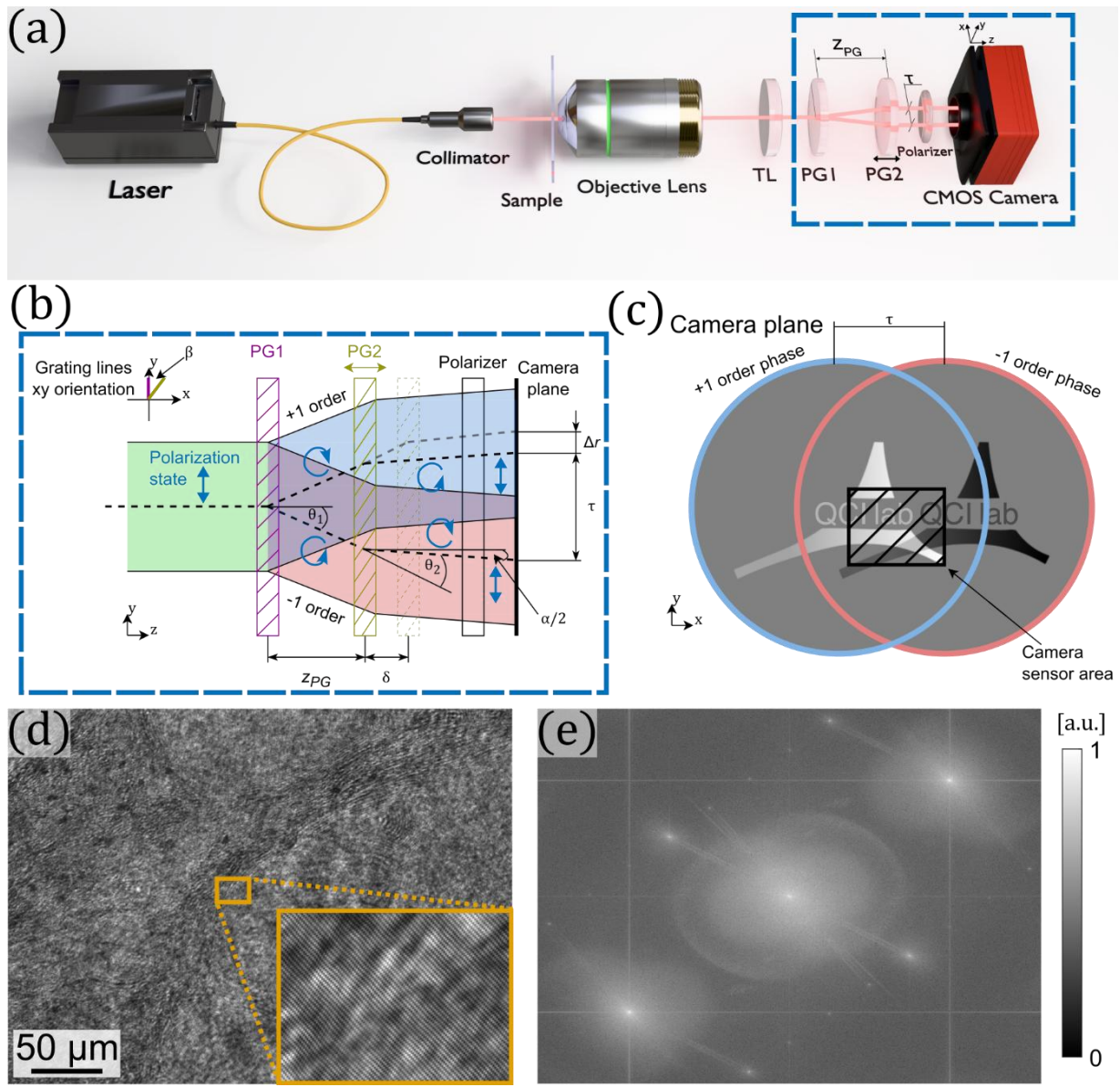

*Fig. 1. (a) Schematic of the polarization grating based common path system. TL – tube lens, PG – polarization gratings, $z_{PG}$ – distance between PG1 and PG2, $\tau$ – shear value between the object beams. (b) Schematic of the shearing module. Blue arrows represent the polarization state (linear or CP of different handedness) after passing through different elements. (c) Visualization of overlapped phase information coming from ±1st diffraction orders. In the area where both beams overlap, the interference pattern is observed – this overlap area is typically larger than camera sensor area. To enable FOV doubling, $\tau$ should be larger than camera sensor dimensions. (d) Interference pattern and (e) its Fourier transform, obtained in our experimental system for human thyroid smear tissue slice.*

# Results

## Reconstruction algorithm

Figure 2 illustrates the processing pipeline of the proposed reconstruction method, whereas Figure 3 shows the phase maps retrieved at each stage of the algorithm.

In our system, the interferometric intensity pattern $I(\mathbf{r})$ recorded by the camera results from the coherent superposition of two complex optical fields corresponding to the +1 and –1 diffraction orders, denoted by $E_{+1}(\mathbf{r})$ and $E_{-1}(\mathbf{r})$, respectively. The corresponding intensity can be expressed as:

$$I(\mathbf{r}) = |E_{+1}(\mathbf{r}) + E_{-1}(\mathbf{r})|^2 = |E_{+1}|^2 + |E_{-1}|^2 + E_{+1}E_{-1}^* + E_{+1}^*E_{-1}, \quad \text{Eq. (1)}$$

where $\mathbf{r} = (x, y)$ denotes the lateral camera coordinates, and * represents complex conjugation.

Each field $E(\mathbf{r})$ can be modeled as a plane wave incident at an angle ($\alpha/2$ in Fig. 1(b)), modulated in both amplitude and phase by the sample:

$$E(\mathbf{r}) = A(\mathbf{r})e^{i(\varphi(\mathbf{r}) + \mathbf{qr})}, \quad \text{Eq. (2)}$$

where $A(\mathbf{r})$ and $\phi(\mathbf{r})$ denote the sample induced amplitude and phase modulation, respectively, and $\mathbf{q}$ ($\mathbf{qr} = q_x x + q_y y$) is a scalar spatial carrier frequency that arises due to the angular tilt $\alpha$ introduced by the PGs (Fig. 1(b)).

Substitution of Eq. (2) into Eq. (1) yields:

$$I = [A_{+1}^2 + A_{-1}^2] + [A_{+1}A_{-1}e^{i(\varphi_{+1} - \varphi_{-1} + 2\mathbf{qr})}] + [A_{+1}A_{-1}e^{i(\varphi_{-1} - \varphi_{+1} - 2\mathbf{qr})}]. \quad \text{Eq. (3)}$$

When $q$ is sufficiently large, the three terms denoted with square brackets in Eq. (3) become spectrally separated in the Fourier domain (Fig. 1(e)) and can be isolated using the Fourier transform method (FTM) [15].

The isolated off-axis interference term yields the following complex-valued field:

$$E_{FTM} = A_{+1}A_{-1}e^{i(\varphi_{+1} - \varphi_{-1})}. \quad \text{Eq. (4)}$$

This retrieved phase, denoted as $\varphi_{\pm 1} = \varphi_{+1} - \varphi_{-1}$, is thus a differential phase between the fields from two diffraction orders. Accurate recovery of $\varphi_{+1}$ and $\varphi_{-1}$ from $\varphi_{\pm 1}$ is only possible in special cases: (1) in total-shear configuration (for large $\tau$) where one of the phases is known or negligible (e.g., for sparse samples), or (2) in gradient phase methods, where the transversal shear between +1 and –1 order phases $\tau$ is very small and $\varphi_{\pm 1}$ is in fact a phase gradient. In more general, total-shear scenario, where both diffraction orders contain overlapping information from different objects, this leads to a replica overlap problem, as visualized in Figs. 3(a)–3(c). In our system, the +1 and –1 diffraction orders correspond to two separate FOVs (with $\tau$ larger than camera diagonal), thus the replicated objects from opposite diffraction order do not fit into the camera sensor. For the purpose of this description, we use the term *replica* to denote the undesired contribution originating from the opposite diffraction order: when reconstructing $\varphi_{+1}$, $\varphi_{-1}$ acts as its replica, and conversely, when reconstructing $\varphi_{-1}$, $\varphi_{+1}$ is considered the replica.

To independently recover $\varphi_{+1}$ and $\varphi_{-1}$, we introduce a novel MRR algorithm. This approach utilizes $N$ (at least two) interferometric patterns $I^{(n)}$, from which the FTM phase reconstructions are obtained ($\varphi_{\pm 1}^{(n)}$). Importantly, each $n^{\text{th}}$ image is acquired with different mutual lateral $(x, y)$ shift of +1 and –1 diffraction order information (Figs. 3(c), 3(d)):

$$\varphi_{\pm 1}^{(n)}(\mathbf{r}) = \varphi_{+1}(\mathbf{r} + n\Delta\mathbf{r}) - \varphi_{-1}(\mathbf{r} - n\Delta\mathbf{r}). \quad \text{Eq. (5)}$$

In our system the $n\Delta\mathbf{r}$ lateral shift value is physically implemented by translating the PG2 by $n\delta$ distance along the optical axis (Fig. 1(b)).

The proposed MRR algorithm begins with $\varphi_{+1}$ and $\varphi_{-1}$ phases estimation (denoted as $\varphi'_{+1}$ and $\varphi'_{-1}$) via averaging and spatial shift compensation of all input $\varphi_{\pm 1}^{(n)}$ phases:

$$\varphi'_{+1}(r) = \frac{1}{N}\sum_{n=1}^{N} \varphi_{\pm1}^{(n)}(r - n\Delta r),\qquad \text{Eq. (6)}$$

$$I\varphi'_{-1}(r) = -\frac{1}{N}\sum_{n=1}^{N} \varphi_{\pm1}^{(n)}(r + n\Delta r).\qquad \text{Eq. (7)}$$

In other words, the FTM retrieved $\varphi_{\pm1}^{(n)}$ phases are firstly transversely shifted so that sought $\varphi_{+1}$ (or $\varphi_{-1}$) phase is fixed in the same transversal area for each $n^{th}$ measurement, while the replica $\varphi_{-1}$ (or $\varphi_{+1}$) is moved by $2n\Delta r$ factor for each measurement. Then the shifted $\varphi_{\pm1}^{(n)}$ phases are pixel-wise averaged resulting with retrieved $\varphi_{+1}$ (or $\varphi_{-1}$) phase contaminated with a motion blurred replica of $\varphi_{-1}$ (or $\varphi_{+1}$). This effect can be also described by substituting Eq. (5) into Eqs. (6)–(7):

$$\varphi'_{+1}(r) = \varphi_{+1}(r) - \frac{1}{N}\sum_{n=1}^{N} \varphi_{-1}(r - 2n\Delta r) = \varphi_{+1}(r) - \xi_{-1}(r),\qquad \text{Eq. (8)}$$

$$\varphi'_{-1}(r) = \varphi_{-1}(r) - \frac{1}{N}\sum_{n=1}^{N} \varphi_{+1}(r + 2n\Delta r) = \varphi_{-1}(r) - \xi_{+1}(r).\qquad \text{Eq. (9)}$$

The above equations state that $\varphi'_{+1}$ and $\varphi'_{-1}$ retrieved via Eqs. (6), (7) contain the desired $\varphi_{+1}$ and $\varphi_{-1}$ phases, which are additionally contaminated with blurred (along $\Delta r$ direction) replicas of opposite order ($\xi_{-1}$ and $\xi_{+1}$ respectively) – Fig. 3(e).

Interestingly, the unknown $\xi_{-1}$ from Eq. (8) may be estimated ($\xi'_{-1}$) by blurring the $\varphi'_{-1}$ from Eq. (7) along $\Delta r$ direction, while unknown $\xi_{+1}$ from Eq. (9) may be estimated ($\xi'_{+1}$) by blurring the $\varphi'_{+1}$ from Eq. (6) along $\Delta r$ direction:

$$\xi'_{-1}(r) = \frac{1}{N}\sum_{n=1}^{N} \varphi'_{-1}(r - 2n\Delta r),\qquad \text{Eq. (10)}$$

$$\xi'_{+1}(r) = \frac{1}{N}\sum_{n=1}^{N} \varphi'_{+1}(r + 2n\Delta r).\qquad \text{Eq. (11)}$$

After substituting the Eqs. (9), (8) into Eqs. (10), (11) respectively we get:

$$\xi'_{-1}(r) = \xi_{-1}(r) - \frac{1}{N}\sum_{n=1}^{N} \xi_{+1}(r - 2n\Delta r) = \xi_{-1}(r) - \psi_{+1}(r),\qquad \text{Eq. (12)}$$

$$\xi'_{+1}(r) = \xi_{+1}(r) - \frac{1}{N}\sum_{n=1}^{N} \xi_{+1}(r + 2n\Delta r) = \xi_{+1}(r) - \psi_{-1}(r).\qquad \text{Eq. (13)}$$

Those equations state that the estimated $\xi'_{-1}$ and $\xi'_{+1}$ blurred replicas contain the sought $\xi_{-1}$ and $\xi_{+1}$, spoiled by the $\psi_{+1}$ and $\psi_{-1}$ factors respectively. Those $\psi$ factors can be interpreted as double blurred versions of $\varphi_{+1}$ and $\varphi_{-1}$ phases respectively. It is to be noted that $\psi_{+1}$ and $\psi_{-1}$ contain information from opposite diffraction order, than the sought $\xi_{-1}$ and $\xi_{+1}$.

Adding $\xi'_{-1}(r)$ and $\xi'_{+1}(r)$ back to the initial estimates (Eqs. (8), (9)) yields refined estimates $\varphi''_{+1}$ and $\varphi''_{-1}$:

$$\varphi''_{+1}(r) = \varphi'_{+1}(r) + \xi'_{-1}(r) = \varphi_{+1}(r) - \psi_{+1}(r),\qquad \text{Eq. (14)}$$

$$\varphi''_{-1}(r) = \varphi'_{-1}(r) + \xi'_{+1}(r) = \varphi_{-1}(r) - \psi_{-1}(r).\qquad \text{Eq. (15)}$$

It is to be noted that after performed operations, the refined $\varphi''_{+1}$ and $\varphi''_{-1}$ phases are completely free from replica factors of opposite diffraction orders (Fig. 3(f)), hence, the replica images are removed analytically. However, they are spoiled by $\psi_{+1}$ and $\psi_{-1}$ factors – of the same diffraction orders as the phase information. The obtained $\varphi''_{+1}$ and $\varphi''_{-1}$ phases may be then treated as a replica-free phase contrast images. However, due to the overlap with $\psi_{+1}$ and $\psi_{-1}$ factors, these phases remain non-quantitative.

To restore quantitative phase information, we apply an iterative correction leveraging the fact that $\varphi$ and $\psi$ in Eqs. (14), (15) are of opposite signs. By making a simple *a priori* assumption: that the object phase has a uniform sign (e.g., $\varphi \geq 0$), the proposed algorithm can iteratively filter out the

$\psi$ information of opposite sign. The correction procedure is summarized in Algorithm 1, shown here for the $\varphi_{+1}$ case:

|   | **Algorithm 1** |
|---|---|
|   | **Inputs:** $\varphi''_{+1}, N, \Delta r, N_{iter}$ <br> **Outputs:** $\varphi^{MRR}_{+1}$ |
| 1 | % **Initial guess** <br> $\varphi^{MRR}_{+1} = \varphi''_{+1}$ |
|   | % **Iterative reconstruction** |
| 2 | **for:** $n_i = 1: N_{iter}$ |
| 3 | % **Filtration – assume that phase is only positive (or only negative)** <br> $\varphi^{MRR}_{+1}(r) = \max(\varphi^{MRR}_{+1}(r), 0)$ |
|   | % **Estimate the $\psi_{+1}$ according to Eq. (9) and Eq. (12)** |
| 4 | $\xi^{MRR}_{+1}(r) = \frac{1}{N}\sum_{n=1}^{N} \varphi^{MRR}_{+1}(r + 2n\Delta r)$ |
| 5 | $\psi^{MRR}_{+1}(r) = \frac{1}{N}\sum_{n=1}^{N} \xi^{MRR}_{+1}(r - 2n\Delta r)$ |
|   | % **Update the $\varphi^{MRR}_{+1}$ according to Eq. (14)** |
| 6 | $\varphi^{MRR}_{+1}(r) = \varphi''_{+1}(r) + \psi^{MRR}_{+1}(r)$ |

The proposed Algorithm 1 begins with the initial guess that the final phase $\varphi^{MRR}_{+1}$ is equal to the $\varphi''_{+1}$ estimated with Eq. (14). Next, the iterative procedure begins. Each iteration starts by applying the previously mentioned *a priori* constraint regarding the sign of the object phase – in this case that object phase must be larger than zero. After this, the $\psi^{MRR}_{+1}$ is estimated from the filtered $\varphi^{MRR}_{+1}$ following the $\varphi_{+1} \to \xi_{+1} \to \psi_{+1}$ relations described in Eq. (9) and Eq. (12). Finally, the new $\varphi^{MRR}_{+1}$ is estimated from the Eq. (14) with the use of estimated $\psi^{MRR}_{+1}$.

The algorithm runs for a user-defined number of iterations $N_{iter}$, after which the quantitative phase result is obtained, Figs. 3(g) and 3(i). As long as the applied *a priori* constraint is valid, the algorithm converges to the true solution, Fig. 3(h). For the experimental data, we ran the algorithm for 10 iterations, as after that we observed no significant improvement in reconstruction accuracy. Apart from $N_{iter}$, the algorithm performance depends also on the number of employed phases $N$ and the shifts $\Delta r$ between subsequent phases. The influence of those parameters is investigated on experimental data in the results section.

It is to be noted, that employed non-negative/non-positive phase constraint should be possible to apply for a vast majority of the potential samples (including biological ones), as typically the measured object has refractive index either higher or lower than the refractive index of its surrounding area. Note, however, that this assumption may fail if different regions of the object exhibit refractive indices both higher and lower than the surrounding medium – in this case the reconstruction would be spoiled with the artifacts. Moreover, the constraint does not apply to high-frequency phase noise (e.g., camera noise or coherent speckle noise) which is typically zero-mean and can take both positive and negative values.

Importantly, while this noise violates the sign constraint, its impact on the reconstruction is typically limited. The initial phase estimation involves averaging steps (Eqs. (6) and (7)), which inherently suppress zero-mean noise components across multiple frames. As a result, the overall noise contribution is attenuated with increasing number of input frames, and its influence onto reconstruction becomes significant only after a large number of iterations. Therefore, in practice, the algorithm remains robust to moderate noise levels. Nevertheless, in scenarios with very low input SNR, reducing the iteration count is advisable to avoid overfitting to noisy features.

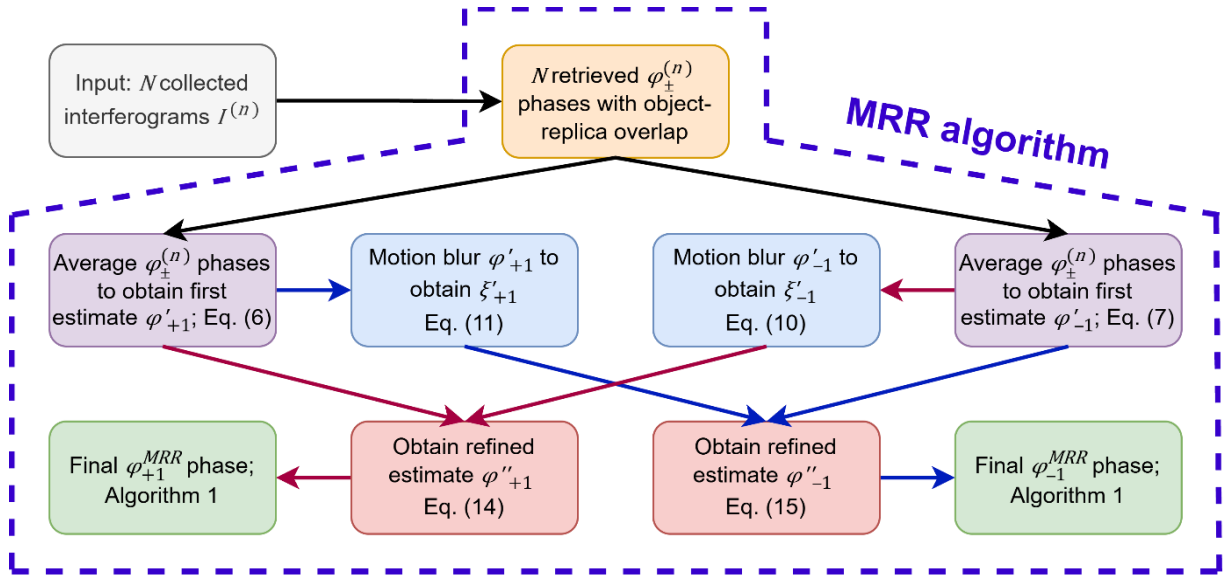

Fig. 2. A schematic of the MRR algorithm that clarifies the processing sequence. The algorithm takes as input N phase images $\varphi_{\pm 1}$, retrieved from interferograms $I^{(n)}$. Those are firstly processed according to Eq. (6) and Eq. (7) to obtain initial estimates $\varphi'_{+1}$ and $\varphi'_{-1}$, spoiled with blurred replica images $\xi_{-1}$ and $\xi_{+1}$ (Eq. (8) and Eq. (9)). Next, the blurred replica images $\xi'_{+1}$ and $\xi'_{-1}$ are estimated with the Eq. (11) and Eq. (10) respectively. Subsequently, the refined phase estimates $\varphi''_{+1}$ and $\varphi''_{-1}$ are retrieved with the use of Eq. (14) and (15). Those refined estimates have analytically removed the overlapped replica, however, they are not quantitative. Finally, the Algorithm 1 algorithm process the refined estimates, to obtain the final, quantitative $\varphi^{MRR}_{+1}$ and $\varphi^{MRR}_{-1}$ phases. Red and blue arrows denote operations required for $\varphi_{+1}$ and $\varphi_{-1}$ phase retrieval respectively.

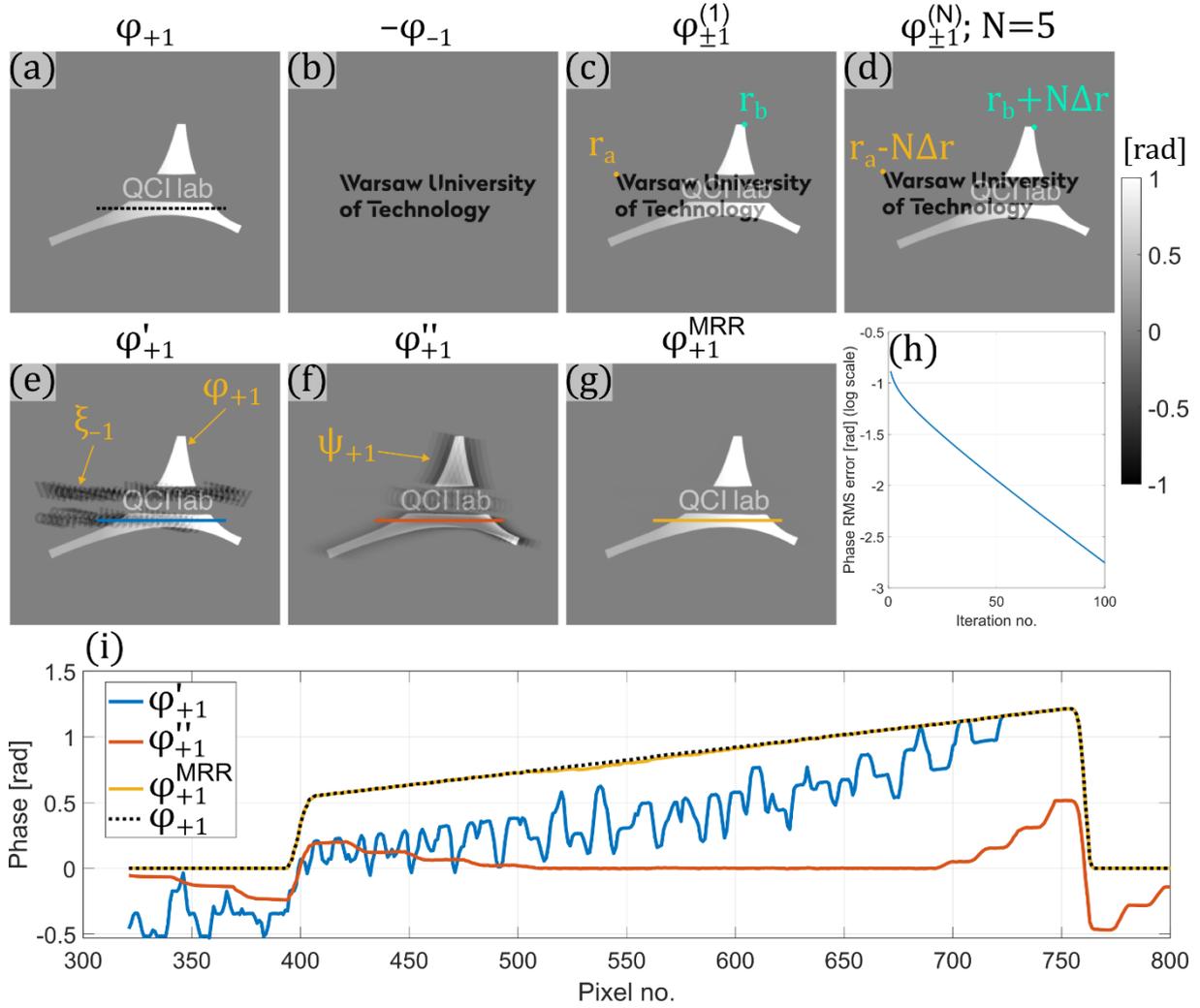

*Fig. 3. Reconstruction results at different stages of the MRR algorithm. (a),(b) Simulated ground-truth $\varphi_{+1}$ and $\varphi_{-1}$ phases. (c), (d) Input $\varphi_{\pm 1}$ phases (for $n=1$ and $n=N$, respectively) used as input to the MRR algorithm, where an overlap between the object and its replica phases is observed. For different values of n, the mutual lateral position between the object and replica changes (compare the location of yellow and green markers in (c), (d)). (e) Initial estimate $\varphi'_{+1}$, showing the desired $\varphi_{+1}$ phase overlaid with a blurred replica $\xi_{-1}$ of the opposite diffraction order. (f) Reconstructed $\varphi''_{+1}$ phase, with the replica analytically removed, but still contaminated by the $\psi_{+1}$ component of the same diffraction order. (g) Final quantitative phase reconstruction $\varphi^{MRR}_{+1}$. (h) Logarithm of the phase RMS error of Algorithm 1 as a function of iteration number, computed as: $Error = \log\left(rms(\varphi_{+1} - \varphi^{MRR}_{+1})\right)$. (i) Cross-sections through phases from (a), (e), (f), and (g), showing the close match of reconstructed $\varphi^{MRR}_{+1}$ and ground truth $\varphi_{+1}$ phases.*

## Experimental results

Typically, interferometric systems require highly coherent illumination to ensure mutual coherence between the interfering beams. However, the use of such sources increases susceptibility to speckle noise and parasitic interference. Interestingly, the proposed system follows a common-path configuration, meaning that both interfering beams propagate through nearly identical optical paths. This significantly minimizes the optical path difference and enables the use of light sources with reduced temporal coherence.

To validate this, we recorded a set of $N = 20$ interference patterns of a custom-designed phase resolution target, with etched elements of 120 nm height (Lyncee Tec, Borofloat 33 glass) using a 5×/0.12 microscope objective. Each interferogram was collected after PG2 grating movement by $\delta = 0.5$ mm, which corresponds to $\Delta r = (5,9)$ pixel shift. Measurements were performed with both a highly coherent laser source (CNI MGL-FN-561, $\lambda = 561$ nm, FWHM $\cong 47$ pm) and a low temporal coherence supercontinuum source with the same wavelength chosen (NKT photonics SuperK EVO, $\lambda = 561$ nm, FWHM $\cong 5$ nm).

Figures 4(a)–4(c) show the results obtained with the coherent laser, while Figs. 4(d)–4(f) correspond to the supercontinuum source. Figures 4(a) and 4(d) show the retrieved $\varphi_{\pm 1}^{(1)}$ raw phase maps, in which the object region (test group "R") overlaps with its replica (test group "H"). The remaining panels present the $\varphi_{+1}^{MRR}$ phase reconstructions obtained using only $N = 3$ frames (Figs. 4(b), 4(e)) or all $N = 20$ frames (Figs. 4(c), 4(f)). As can be observed, the reconstructed test patterns from group "R" are free from replica artifacts in all $\varphi_{+1}^{MRR}$ results, confirming that proposed algorithm performed successfully.

Inserts in Fig. 4 highlight a single element from group "R" consisting of 3 µm-wide lines, which are clearly resolved in all $\varphi_{+1}^{MRR}$ reconstructions. This aligns with the theoretical resolution of the system ($\lambda/2\text{NA} \approx 2.3$ µm), indicating that the system reaches diffraction limited resolution for both light sources. Notably, the supercontinuum source yields reconstructions with substantially reduced coherent noise, visible in the cross-section plots and also shown quantitatively as the standard deviation ($\sigma$) of object-free area (marked with green color in Fig. 4). Moreover, an increase in the number of input frames reduces phase noise further, as clearly seen in the laser-based results, which initially exhibit higher noise levels. This improvement is attributed to the averaging process in Eqs. (6) and (7), which also reduces the sample uncorrelated noise present in the individual $\varphi_{\pm 1}^{(1)}$ phase maps following the $\sigma_{1:N} = \frac{1}{\sqrt{N}}\sigma_n$ equation [60]. This suggests that, from an SNR perspective, low-coherence illumination enables high-quality reconstructions with fewer input frames, offering the potential for improved temporal resolution in dynamic measurements.

Finally, one should note that the reconstructed images in Figs. 4(b), 4(e) and especially Figs. 4(c), 4(f) appear slightly larger than the original $\varphi_{\pm 1}^{(1)}$ phase maps shown in Figs. 4(a), 4(d). This enlargement is not due to poor figure formatting, but results from the use of shifted phase measurements (with displacements of $+n\Delta \boldsymbol{r}$), which effectively expand the reconstructed field of view by $N\Delta \boldsymbol{r}$ pixels. In subsequent figures, we present the $\varphi_{+1}^{MRR}$ results cropped to match the original $\varphi_{\pm 1}^{(1)}$ field of view for consistency.

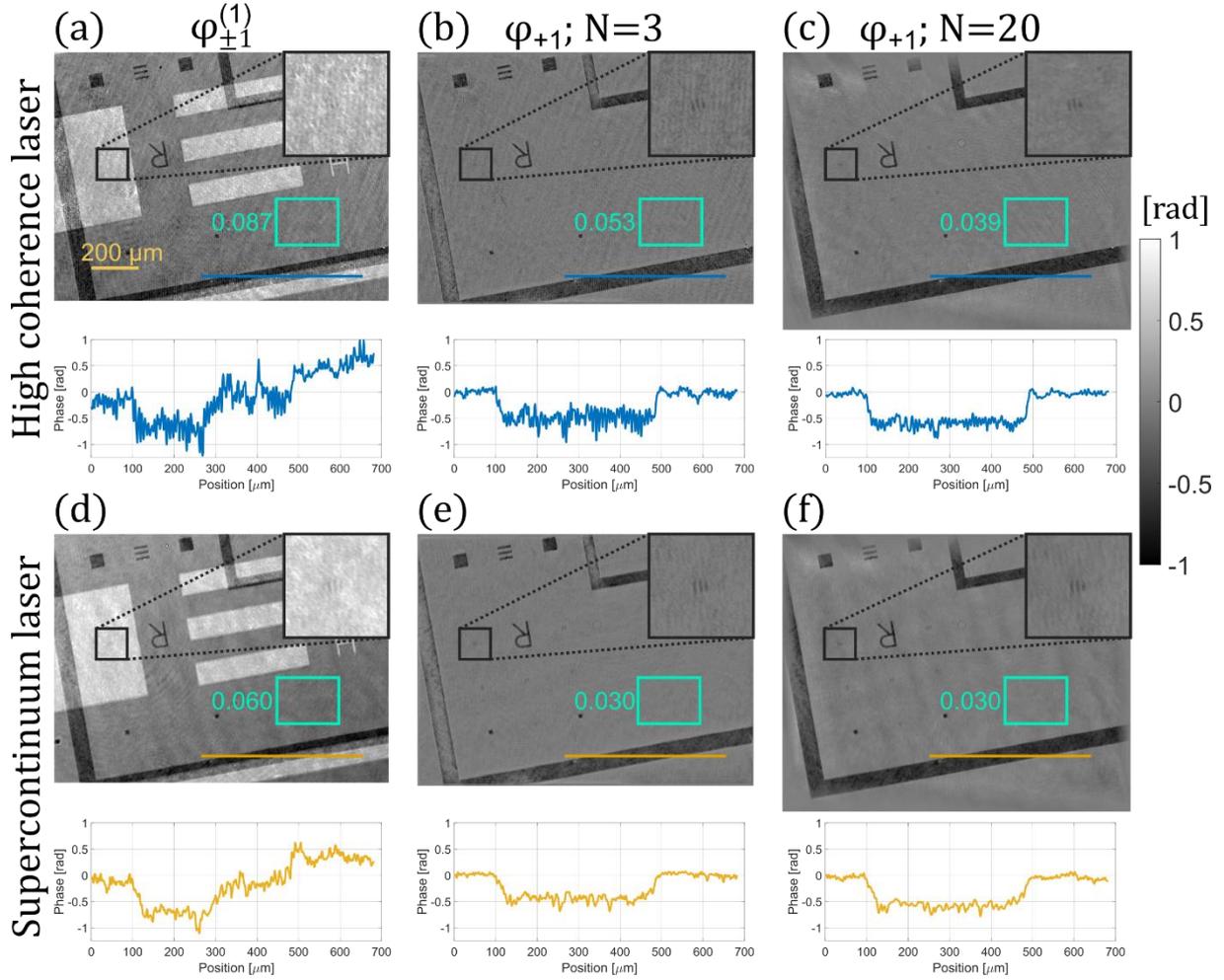

Fig. 4. Phase imaging results of phase resolution target. (a)-(c) input $\varphi_{\pm1}^{(n)}$ image and $\varphi_{+1}^{MRR}$ reconstructions for $N = 3$ and $N = 20$ respectively for high coherence laser illumination. (d)-(f) input $\varphi_{\pm1}^{(n)}$ image and $\varphi_{+1}^{MRR}$ reconstructions for $N = 3$ and $N = 20$ respectively for low temporal coherence supercontinuum illumination. Numerical values shown in green indicate the phase standard deviation (in radians) calculated within the object-free regions marked by green rectangles.

One of the limitations of the proposed method lies in its ability to recover low spatial frequency components of the object phase. When the lateral size of the measured object is relatively large compared to the total effective shift $N\Delta\boldsymbol{r}$, the low-frequency information of the factor $\psi$ becomes nearly indistinguishable from that of the phase $\varphi$ in Eqs. (14) and (15), resulting in the mutual cancellation of these components.

In theory, such missing information can still be recovered iteratively using Algorithm 1 procedure. However, this may require a large number of iterations, which increases sensitivity to noise and can degrade the reconstruction quality. To evaluate the method's performance in retrieving low spatial frequencies, we recorded a series of 200 interferograms of a large square test element from group "G" (500 × 500 μm) overlapped with smaller features from group "N". Measurements were conducted using a 5×/0.12 microscope objective and a supercontinuum laser source. Each interferogram was acquired after translating the PG2 along the optical axis by $\delta = 0.05$ mm, corresponding to a sub-pixel displacement of $\Delta\boldsymbol{r} = (0.6, 0.7)$.

We then performed a set of reconstructions for different values of the number of images $N$ (ranging from 2 to 20) and by choosing different grating shift steps $\delta$ between subsequent frames (ranging from 0.05 mm to 0.5 mm). Figure 5 summarizes the results of this experiment. Figure 5(a) shows the retrieved $\varphi_{\pm 1}^{(n)}$ phase maps for $n = 1$, $n = 100$, and $n = 200$, while Fig. 5(b) presents the replica phase image reconstructed using $N = 20$ interferograms with $\delta = 0.5$ mm. Figures 5(c) and 5(d) display reconstructed phase maps of the "G" group element for various combinations of $N$ and $\delta$. As both parameters increase, the reconstructed center shows fewer voids and near-zero artifacts – confirming improved retrieval of low-frequency content.

To quantitatively assess the recovery of low spatial frequencies, we calculated the average phase value within the green region marked in Fig. 5(c4) for all tested configurations. As a reference, we used a value of -0.55 radians – a mean value obtained from a replica-free area marked with a green rectangle in Fig. 5(a3). We then computed the absolute deviation of the reconstructed phase from this reference value across all tested $(N, \delta)$ combinations.

The results, summarized in Fig. 5(e), reveal that the reconstructed phase was underestimated, ranging from a deviation of 0.52 radians for $N = 2$ and $\delta = 0.05$ mm to just 0.13 radians for $N = 20$ and $\delta = 0.5$ mm. These findings confirm that increasing both the number of interferometric frames and the shift amplitude significantly improves the recovery of low spatial frequency phase content. For the largest tested $N$ and $\delta$ values, the effective $N\Delta r$ shift corresponded to 120 × 140 px, which is approximately 5–6 times smaller than the dimensions of the measured element (725 × 725 px). This suggests that, in order to achieve quantitative retrieval of low-frequency phase information, the $N\Delta r$ value should be at least larger than the one fifth of lateral size of the object.

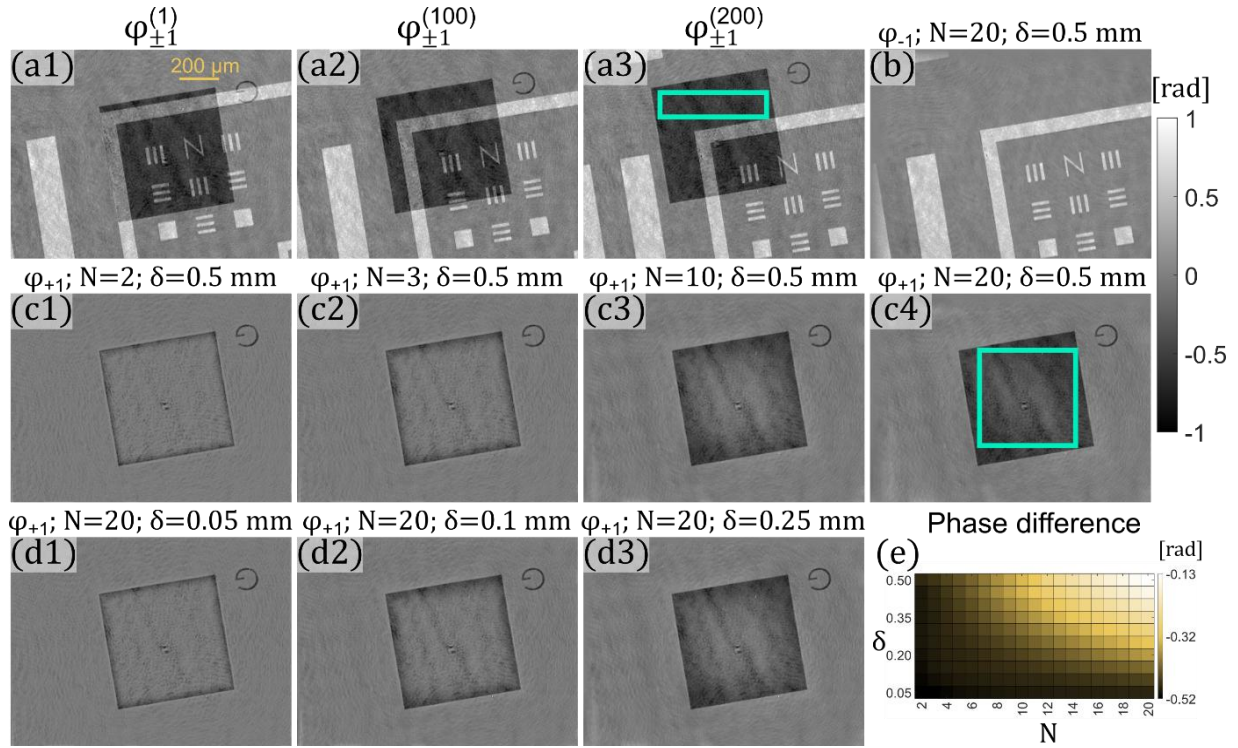

*Fig. 5. Phase imaging results of large dimensions square element (500x500 µm) from phase resolution target. (a) FTM retrieved phases with visible replica overlap for $n = 1$, $n = 100$ and $n = 200$. (b) reconstructed $\varphi_{-1}^{MRR}$. (c),(d) reconstructed $\varphi_{+1}^{MRR}$ for different $N$ and $\delta$ values. (e) Difference between the element average phase from reference measurement (marked with green area in (a3)) and the average phase from reconstruction (marked with green area in (c4)), for different N and $\delta$ combinations.*

Figure 6 presents an evaluation of the algorithm's performance for a fixed biological sample – *Saccharomyces cerevisiae* yeast cells. The dataset was acquired using a 20×/0.75 microscope objective, under supercontinuum laser illumination, with a diffraction grating shift of $\delta = 0.5$ mm between subsequent acquisitions. Relatively low magnification and doubled FOV predestine these R2D-QPI setup for high-throughput imaging of widefield biological specimens.

Fig. 6(a) displays the retrieved $\varphi_{\pm 1}^{(n)}$ phases for $n = 1$ and $n = 20$, demonstrating strong interference between the object and its replica due to the dense nature of the sample. Figs. 6(b)–6(d) show the reconstructed $\varphi_{+1}^{MRR}$ and $\varphi_{-1}^{MRR}$ phases for the +1 and –1 diffraction orders FOVs using $N = 2, 3$, and 20 input images.

The yeast sample is highly confluent, making it practically impossible to isolate regions where the object and replica fields do not overlap. As a result, both $\varphi_{\pm 1}^{(1)}$ and $\varphi_{\pm 1}^{(20)}$ phase images (Fig. 6(a)) exhibit significant distortions due to the mutual overlap of the object and replica FOVs.

Despite this, the proposed MRR algorithm successfully separates the object and replica contributions, enabling simultaneous reconstruction of both diffraction orders' FOVs. Notably, the relatively small dimensions of individual yeast cells (approximately 25 px ≈ 4.3 μm in diameter) imply that most of the sample's phase content is concentrated in high spatial frequencies. Consequently, even reconstructions based on as few as $N = 2$ or $N = 3$ input images closely match the full $N = 20$ reconstruction, as evidenced by the zoomed-in areas in Figs. 6(b)–6(d).

However, in the case of large yeast clusters, reconstructions with lower $N$ tend to underestimate phase values in the central regions of the clusters. This observation suggests that for isolated or small-sized biological features, accurate phase retrieval can be achieved using only a minimal number of measurements. In contrast, larger structures require increased values of $N$ to avoid phase loss, particularly in the low spatial frequency components.

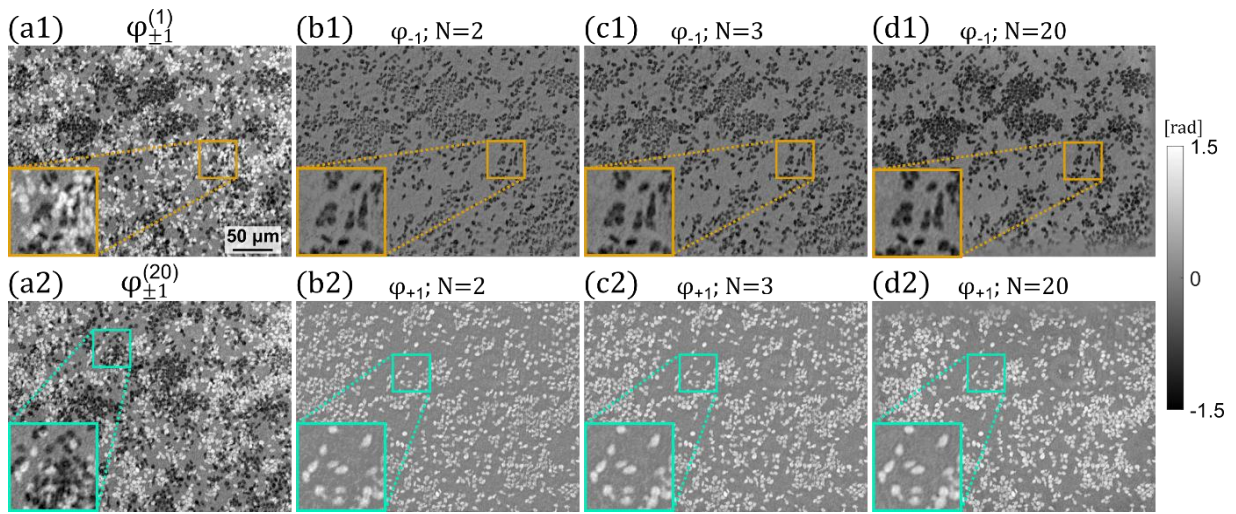

Fig. 6. Phase imaging results of Saccharomyces cerevisiae yeast cells. (a) FTM retrieved phases with visible replica overlap for $n = 1$ and $n = 20$. (b)-(d) reconstructed $\varphi^{MRR}$ phases for $N = 2$, $N = 3$ and $N = 20$ input images respectively, for both +1 and –1 order FOVs. Zoomed in area in (a2) is laterally shifted comparing to images (b2)-(d2) by $N\Delta r$ value (due to the PG2 movement by $N\delta$), in order to show the same $\varphi_{+1}$ yeasts.

Figure 7 presents the results of phase reconstruction for a human thyroid smear tissue. The dataset was acquired with both 5×/0.12 (for larger FOV) and 20×/0.75 (for higher resolution) microscope objectives, using a supercontinuum laser and grating shifts of $\delta = 0.5$ mm between interferograms. In total, $N = 20$ interferometric patterns were recorded for each objective.

Fig. 7(a) shows the retrieved $\varphi_{\pm 1}^{(1)}$ phase images for the both employed objectives. The regions of interest were chosen to align the –1 order FOV between 5× and 20× acquisitions (highlighted with green rectangle in Fig. 7(b)). However, due to the different magnification, the +1 diffraction order FOVs do not overlap between the two objectives, as shown in Fig. 7(c).

Despite the complexity of the biological tissue, the proposed MRR algorithm successfully separates the overlapping phase contributions for both diffraction orders and both magnifications. This demonstrates the robustness of the method, even in the presence of densely packed and irregular phase structures typically encountered in tissue samples.

Moreover, the consistency of the reconstructed structures across different objectives highlights the scalability and versatility of the method. While the 5× objective allows for rapid imaging over larger tissue areas, especially with FOV doubling effect, the 20× reconstruction provides higher spatial resolution. This makes the approach well-suited for multi-scale imaging workflows, where both overview and high-resolution inspection are required within the same optical setup.

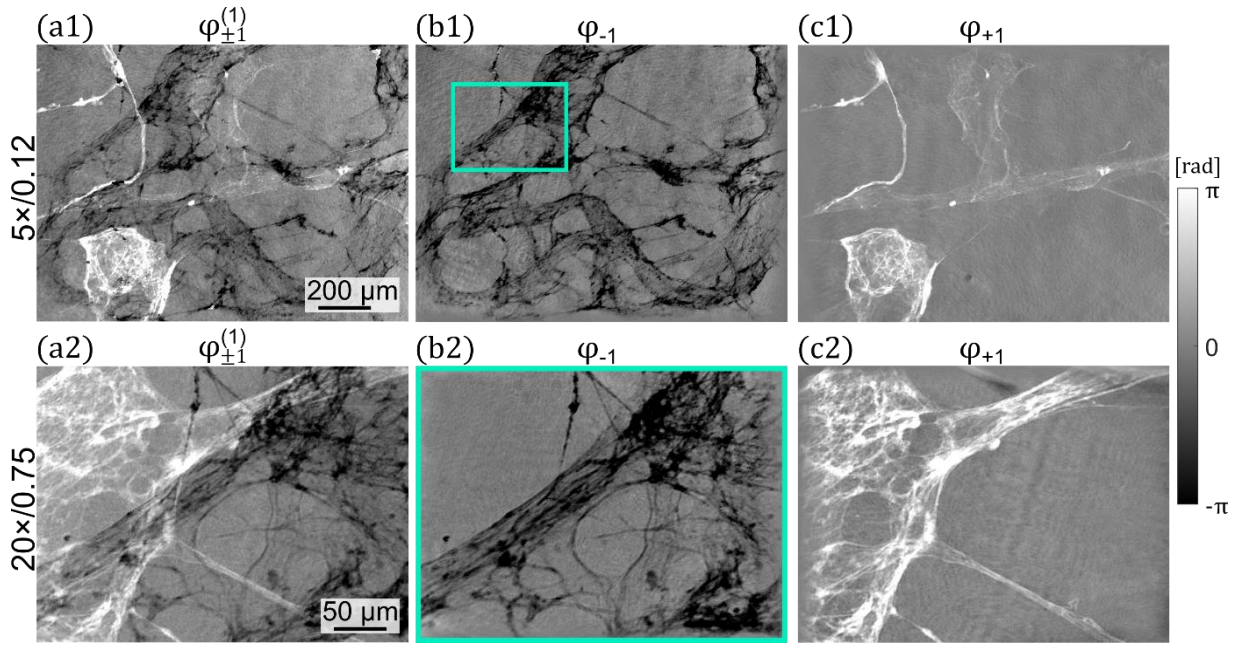

Fig. 7. Phase imaging results of human thyroid smear tissue with thyrocytes and colloid visible. (a) FTM retrieved phases with visible replica. (b), (c) reconstructed $\varphi^{MRR}$ phases +1 and –1 order FOVs respectively. Top – images for 5×/0.12 objective. Bottom – images for 20×0.75 objective. 20× $\varphi_{-1}^{MRR}$ FOV was chosen so that to match 5× $\varphi_{-1}^{MRR}$ FOV, as marked with green rectangles.

# Discussion and conclusions

In this work, we introduced a general shear-scanning common-path interferometric R2D-QPI approach, in which controlled variations of the lateral shear enable separation of overlapped diffraction-order fields of view. To achieve this, we developed the novel MRR algorithm that reconstructs the true object phase from as few as two measurements with different shear values. While in this study the method is demonstrated using a polarization grating-based interferometer, the concept is not limited to this architecture and can be directly applied to other common-path shearing configurations with tunable shear (e.g., grating-, Michelson-, Sagnac-, Wollaston-, beam displacer- or shearing plate-based systems) [43].

The proposed method has proven to be both robust and accurate. It allows complete and analytical removal of unwanted replica contributions from as few as two input images, even for extremely dense and highly confluent samples. Importantly, while two images are sufficient for complete separation, increasing the number of acquired interferograms further improves the reconstruction by averaging out noise, enhancing the signal-to-noise ratio, and providing more accurate recovery of low-spatial-frequency phase components.

Nevertheless, two main limitations remain. First, the method requires mechanical scanning of the diffraction grating, which reduces temporal resolution and may be unfavorable in applications requiring high stability and repeatability. This limitation could be potentially mitigated by multiplexed detection (using multiple cameras to simultaneously capture interferograms at different shear values), or by replacing mechanical scanning with wavelength tuning [61], which naturally results in shear variation. Second, the algorithm shows reduced accuracy in retrieving low-frequency phase components. As demonstrated, this limitation can be mitigated by increasing the number of acquired interferograms or by enlarging the shear step between them.

In summary, the proposed MRR approach provides a powerful and versatile solution to one of the central challenges of common-path interferometry – the separation of object and replica phases. It enables accurate reconstruction from a minimal number of input images, even in dense and complex samples, while simultaneously offering the benefit of an expanded FOV. These unique advantages make the method particularly attractive for high-throughput biological imaging, tissue diagnostics, and quantitative phase analysis of confluent cell cultures. Future directions include eliminating the need for mechanical scanning, improving retrieval of low-frequency phase features, and extending the approach to multimodal or real-time implementations. Together, these developments could establish MRR as a broadly applicable tool in next-generation label-free imaging.

# Acknowledgements

Research was funded by the project no. WPC3/2022/47/INTENCITY/2024 funded by the National Centre for Research and Development (NCBR) under the 3rd Polish-Chinese/Chinese-Polish Joint Research Call (2022). The research was conducted on devices cofounded by the Warsaw University of Technology within the Excellence Initiative: Research University (IDUB) programme. This work was supported by the National Natural Science Foundation of China (62227818, 62361136588, 62575139), National Key Research and Development Program of China (2024YFE0101300)